\documentclass{article}
\pdfoutput=1
\usepackage{ijcai15}

\usepackage{times}
\usepackage{helvet}
\usepackage{courier}
\usepackage{graphicx}
\usepackage{subfigure}
\usepackage{url}
\usepackage{algorithm,algorithmic}
\usepackage{multirow}

\begin{document}
\title{A Picture Tells a Thousand Words -- About You! \\User Interest Profiling from User Generated Visual Content}
%\author{ Submitted for Blind Review}

\author{Quanzeng You \and Jiebo Luo \\
Department of Computer Science \\
University of Rochester \\
Rochester, NY 14623 \\
\{qyou, jluo\}@cs.rochester.edu \\
\And Sumit Bhatia \\
IBM Almaden Research Centre \\
650 Harry Rd, San Jose, CA 95120\\
\{sumit.bhatia\}@us.ibm.com  \\
}

\maketitle
\begin{abstract}
Inference of online social network users' attributes and interests has been an active research topic. Accurate identification of users' attributes and interests is crucial for improving the performance of personalization and recommender systems. Most of the existing works have focused on textual content generated by the users and have successfully used it for predicting users' interests and other identifying attributes. However, little attention has been paid to user generated visual content (images) that is becoming increasingly popular and pervasive in recent times. We posit that images posted by users on online social networks are a reflection of topics they are interested in and propose an approach to infer user attributes from images posted by them. We analyze the content of individual images and then aggregate the image-level knowledge to infer user-level interest distribution. We employ image-level similarity to propagate the label information between images, as well as utilize the image category information derived from the user created organization structure to further propagate the category-level knowledge for all images. A real life social network dataset created from Pinterest is used for evaluation and the experimental results demonstrate the effectiveness of our proposed approach.
\end{abstract}

\section{Introduction}
\label{intro}
Online Social Networks (OSNs) such as Facebook, Twitter, Pinterest, Instagram, etc. have become a part and parcel of modern lifestyle. A study by Pew Research centre\footnote{http://www.pewinternet.org/2013/12/30/social-media-update-2013/} reveals  that three out of every four adult internet users use at least one social networking site. Such large scale adoption of OSNs and active participation of users have led to research efforts studying relationship between users' digital behavior and their demographic attributes (such as age, interests, and preferences)  that are of particular interest to social science, psychology, and marketing. A large scale study of about 58,000 Facebook users performed by Kosinski et al.~\shortcite{kosinski2013private} reveals that digital records of human activity can be used to accurately predict a range of personal attributes such as age, gender, sexual orientation, political orientation, etc. Likewise, there have been numerous works that study variations in language used in social media with age, gender, personality, etc.~\cite{burger2011discriminating,gender_lexical_variation,facebook_plos_one}.  While most of the popular OSNs studied in literature are mostly text based, some of them (e.g., Facebook, Twitter) also allow people to post images and videos. Recently, OSNs such as Instagram and Pinterest that are majorly image based have gained popularity with almost 20 billion photos already been shared on Instagram and an average of 60 million photos being shared daily\footnote{http://instagram.com/press/}.

\begin{figure}
\centering
\includegraphics[width=.47\textwidth]{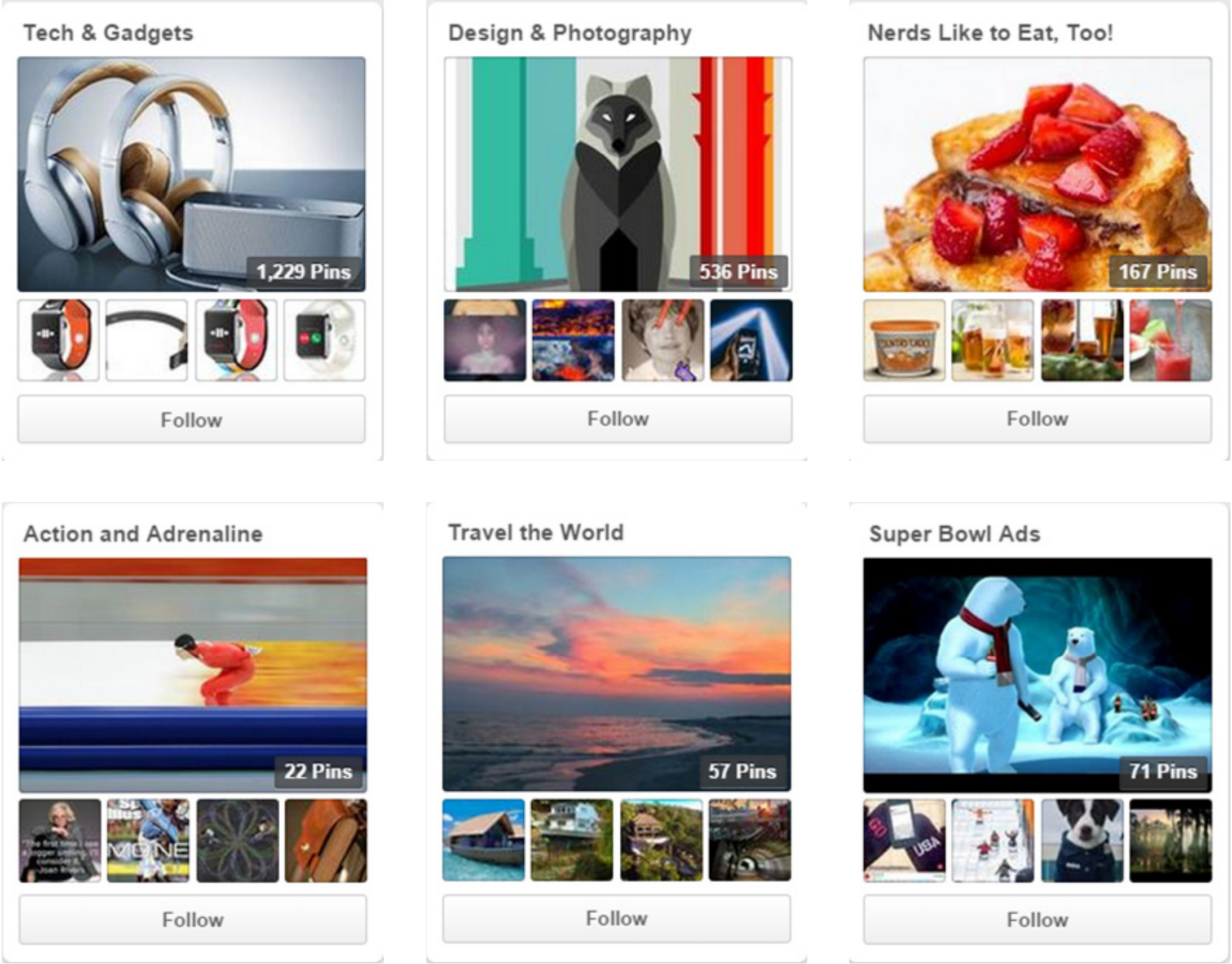}
\caption{Example pinboards from one typical Pinterest user.}
\label{fig:example:pinboards}
%\vspace{-10pt}
\end{figure}
The most appealing aspect of image based OSNs is that visual content is \emph{universal} in nature and thus, not restricted by the \emph{barriers of language}. Users from different cultural backgrounds, nationalites, and speaking different languages can easily use the same \emph{visual language} to express their feelings.  Hence, analyzing the content of user posted images is an appealing idea with diverse applications. Some recent research efforts also provide support for the hypothesis that images posted by users on OSNs may prove to be useful for learning various personal and social attributes of users. Lovato et al.~\shortcite{lovato2013we} proposed a method to learn users' latent visual preferences by extracting aesthetics and visual features from images favorited by users on Flickr. The learned models can be used to predict images likely to be favorited by the user on Flickr with reasonable accuracy. Cristani et al.~\shortcite{cristani2013unveiling} infer personalities of users by extracting visual patterns and features from images marked as favorites by users on Flickr. Can et al.~\shortcite{predicting_retweet_count} utilize the visual cues of tweeted images in addition to textual and structure-based features to predict the retweet count of the posted image. Motivated by these works,  we investigate \emph{if the images posted by users on online social networks can be used to predict their fine-grained interests or preferences about different topics}. To understand this better, \figurename~\ref{fig:example:pinboards} shows several randomly selected pinboards (collection of images, as they are called in Pinterest) for a typical Pinterest user as an example. We observe that different \emph{pins} (a pin corresponds to an image in Pinterest) and pinboards are indicative of the user's interests in different topics such as sports, art and food. We posit that the visual content of the images posted by a user in an OSN is a reflection of her interests and preferences. Therefore, an analysis of such posted images can be used to create an interest profile of the user by analyzing the content of individual images posted by the user and then aggregating the image-level knowledge to infer user-level preference distribution at a fine-grained level.

%Sumit's edit
\subsection{Problem Formulation and Overview of Proposed Approach}
\textbf{Problem Statement:} Given a set $\mathcal{I}$ of images posted by the user \emph{u} on an OSN, and a set $\mathcal{C}$ of interest categories, output a probability distribution over categories in $\mathcal{C}$ as the interest distribution for the user. In order to solve this problem, we first need to understand the relationships between different categories (topics) and underlying characteristics (or features) of user posted images in the training phase. These learned relationships can then be used to predict distribution over different interest categories by analyzing images posted by a new user.  Even though state-of-the-art machine learning algorithms, especially developments in deep learning, have achieved significant results for individual image classification~\cite{krizhevsky2012imagenet}, we believe that  incorporating OSN and user specific information can provide further performance gains. Different image based OSNs offer users capability to group together similar images in the form of albums/pinboards, etc. In Pinterest, users create pinboards for a given topic and collect similar images in the pinboard. Given this human created group information, it is reasonable to assume that strong correlations exist between objects belonging to the same curated group or categorization. For example, a user may have two pinboards belonging to the \emph{Sports} category, one for images related to soccer and one for images related to basketball. Therefore, even though all images in the two pinboards will share some common characteristics, images within each pinboard will share some additional similarities. Motivated by these observations, we employ image level and group level label propagation in order to build more accurate learning models. We employ image-level similarity to propagate the label information between images and category correlations are employed to further propagate the category-level knowledge for all images.

\section{Related Work}
%In this section, we review works which are closely related to our motivation for online user profiling. Specifically, we focus on works in user profiling and visual content analysis.
\subsection{User Profiling from Online Social Networks}
It is discovered that in social network people's relationship follows the rule \textit{birds of a feather flock together}~\cite{mcpherson2001birds}. Similarly, people in online social network also exhibit such kind of patterns. Online social network users connected with other users may be due to very different reasons. For instance, they may try to rebuild their real world social networks on the online social network. However, most of the time, people hope to show part of themselves to the rest of the world. In this case, the content generated by online social network users may help us to infer their characteristics. Therefore, we are able to build more accurate and more targeted systems for prediction and recommendation.

There are many related works on profiling different online users' attributes. Location is quite important for advertisement targeting, local customer personalization and many novel location based applications. In Cheng et al.~\shortcite{cheng2010you}, the authors proposed an algorithm to estimate a city level location for Twitter users. More recently, Li et al.~\shortcite{li2012towards} proposed a unified discriminative influence model to estimate the home location of online social users. They unified both the social network and user-centric signals into a probabilistic framework. In this way, they are able to more accurately estimate the home locations. Meanwhile, since most social network users tend to have multiple social network accounts, where they exhibit different behaviors on these platforms. Reza and Huan~\shortcite{zafarani2013connecting} proposed MOBIUS for finding a mapping of social network users across different social platforms according to people's behavioral patterns. The work in Mislove et al.~\shortcite{mislove2010you} also proposed an approach trying to inferring user profiles by employing the social network graph. In a recent work, Kosinski et al.~\shortcite{kosinski2013private} revealed the power of using Facebook to predict private traits and attributes for Facebook volunteer users. Their results indicated that simple human social network behaviors are able to predict a wide range of human attributes, including sexual orientation, ethnic origin, political views, religion, intelligence and so on. Li et al.~\shortcite{li2014user} defined \textit{discriminative correlation} between attributes and social connections, where they tried to infer different attributes from different circles of relations.
\subsection{Visual Content Analysis}
Visual content becomes increasingly popular in all online social networks. Recent research works have indicated the possibility of using online user generated visual content to learn personal attributes. In Kosinsky et al.~\shortcite{kosinski2013private}, a total of $58,000$ volunteers provided their Facebook likes as well as detailed demographic profiles.  Their results suggest that digital records of human activities can be used to accurately predict a range of personal attributes such as age, gender, sexual orientation, and political orientation. Meanwhile, the work in Lovato et al.~\shortcite{accv_personality} tried to build the connection between cognitive effects and the consumption of multimedia content. Image features, including both aesthetics and content, are employed to predict online users' personality traits. The findings may suggest new opportunities for both multimedia technologies and cognitive science. More recently, Lovato et al.~\shortcite{ieee_biometrics} proposed to learn users' biometrics from their collections of favorite images. Various perceptual and content visual features are proven to be effective in terms of predicting users' preference of images. You, Bhatia and Luo~\shortcite{you2014} exploited visual features to determine the gender of online users from their posted images.

\section{Proposed Approach}
As discussed in Section~\ref{intro}, the problem of inferring user interests can be considered as an image classification problem. However, in contrast to the classical image classification problem where the objective is to maximize classification performance at individual level, we are focused more on learning the overall user-level \emph{image category distribution}, which in turn yields users' interests distribution. In our work, we use data crawled from pinterest.com, which is one of the most popular image based social networks. In Pinterest, users can share/save images that are known as \emph{pins}. Users can categorize these pins into different \emph{pinboards} such that a pinboard is collection of related pins (images).  Also, while creating a pinboard the user has to chose a descriptive category label for the pinboard from a pre-defined list specified by Pinterest. There are a total of 34 available categories for users to chose from (listed in \tablename~\ref{tab:pin:cat}). A typical Pinterest manages/owns many different pinboards belonging to different categories and each pinboard will contain closely related images of the same category. Pinterest users mainly use pinboards to attract other users and also to organize the images of interest for themselves. Therefore, often times they will choose interesting and high-quality pins to add to their pinboards and one can use these carefully selected and well organized high quality images to  infer the interests of the user.

\begin{table*}[!thb]
    \caption{List of 34 categories in Pinterest.}
    \label{tab:pin:cat}
    \centering
    \small
    \begin{tabular}{*{7}{|l}|}
    \hline
    Animals &  Architecture &  Art &  Cars \& Motorcycles &  Celebrities &  Design &  DIY \& Crafts\\ \hline
    Education &  Film, Music \& Books &  Food \& Drink &  Gardening &  Geek &  Hair \& Beauty &  Health \& Fitness\\ \hline
    History &  Holidays \& Events &  Home Decor &  Humor &  Illustrations \& Posters &  Kids &  Men's Fashion\\ \hline
    Outdoors &  Photography &  Products &  Quotes &  Science \& Nature &  Sports &  Tattoos\\ \hline
    Technology &  Travel &  Weddings &  Women's Fashion &  Other &  &  \\
    \hline
    \end{tabular}
\end{table*}
\begin{figure*}
\centering
\includegraphics[width=.8\textwidth]{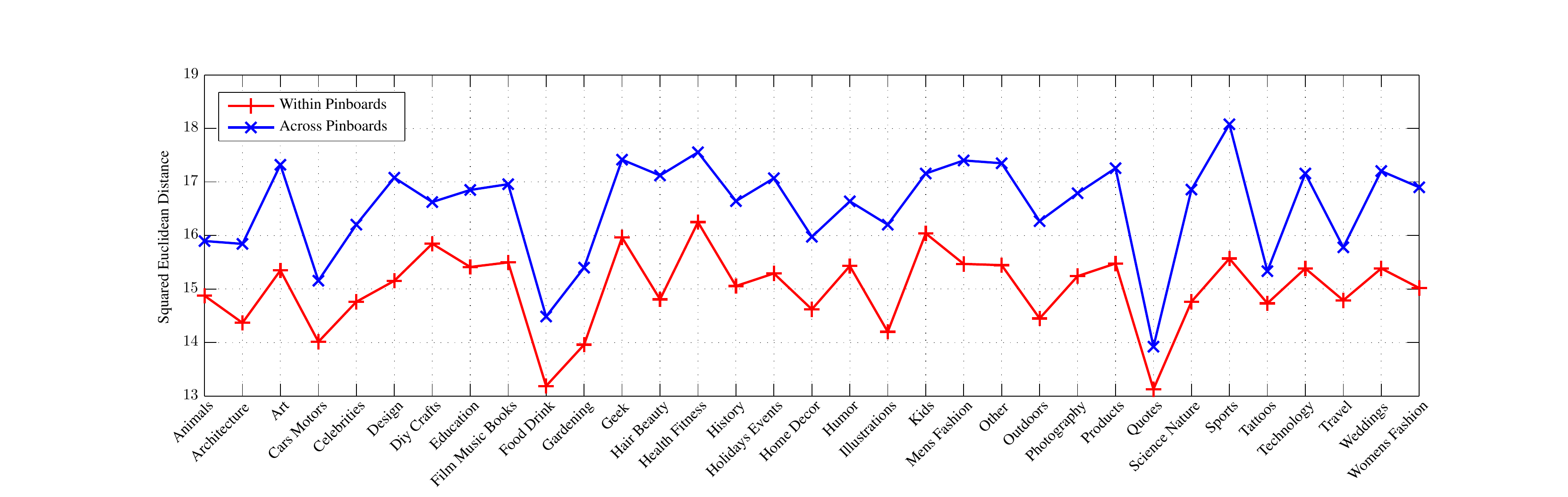}
\caption{Average distance between images in different groups.}
\label{fig:pin:dist}
\vspace{-10pt}
\end{figure*}
\subsection{Training Image-level Classification Model}
We employ the Convolutional Neural Network (CNN) to train our image classifier. The main architecture comes from the Convolutional Neural Network proposed by Krizhevsky et al.~\shortcite{krizhevsky2012imagenet}, that has achieved state of the art performance in the challenging ImageNet classification problem. We train our model on top of the model of Krizhevsky et al.~\shortcite{krizhevsky2012imagenet}). For a detailed description of the architecture of the model, we direct the reader to the original paper. In particular, we fine-tune our model by employing images and labels from Pinterest. For each image, we assign the label of the image to be the same as the label of the pinboard it belongs to. More details on the preparation of training data for our model will be discussed in the experimental section.
%We employ the recently proposed Convolutional Neural Network (CNN) to train our image classifier model. The main architecture comes from the Convolutional Neural Network proposed in~\cite{krizhevsky2012imagenet}, which has achieved the state of the art performance in the challenging ImageNet classification problem. \figurename~\ref{fig:dl:group} shows the CNN architecture, where the number indicates the size and number of filters in each layer. We train our model on top of the model in~\cite{krizhevsky2012imagenet}. In particular, we fine-tune our model by employing images and labels from Pinterest. For each image, we assign the label of the image to be the same with the label of the pinboard it belongs to. More details on the preparation of training our model will be discussed in the experimental section.
%\begin{figure}
%\includegraphics[width=0.45\textwidth]{./figure/dl}
%\caption{Convolutional Neural Network with Within Group Constraint.}
%\label{fig:dl:group}
%\end{figure}
From the trained deep CNN model, it is possible to extract deep features for each image. These features, also known as high-level features, have shown to be more appropriate for performing various other image related tasks, such as image content analysis and semantic analysis~\cite{bengio2009learning}. Specifically, we extract the deep CNN features from the last fully connected layer of our trained CNN model and employ these deep features for label propagation as described in the next section. It is noteworthy that this work differs from image annotation, which deals with individual images and have been extensively studied.  As will become more clear, the approach we take in the form of label propagation, is designed to exploit the strong connection between the images
curated within and across collections by users. For the same reason, the typical noise in user data is suppressed in Pinterest due the same user curation process.

\subsection{Image and Group Level Label Propagation for Prediction}
During the prediction stage, we try to solve a more general problem, where we are given a collection of images without group information. However, we want to predict the users' interests from these unorganized collection of images. We propose to use label propagation for image labels. The work in Yu et al.~\shortcite{yu2011collection}) also tried to use collection information for label propagation. However, differently from their work, where collection information is employed to extract sparsity constraint to the original label propagation algorithm, we propose to impose the additional group-level similarity to further propagate the image labels for the same user.

Meanwhile, we observe that for most of the categories in~\tablename~\ref{tab:pin:cat}, the average distance between images in the same pinboard have closer distance than images in the same category. \figurename~\ref{fig:pin:dist} shows the average distance between images in the same pinboards and the average distance between images in different pinboards but in the same categories. The distance is represented by the squared Euclidean distance between deep features from last layer of Convolutional Neural Network. Intuitively, this can be explained by the fact that most of the categories are quite diverse in that they can contain quite different sub-categories. On the other hand, users generally create pinboards to collect similar images into the same group. These images are more similar in that they are more likely to be in the same sub-category of the category label chosen by the user. Hence, the motivation for implementing label propagation at user level.

Let $n$ be the number of categories, we define matrix $G\in R^{n\times n}$ to be the affinity matrix between these $n$ categories ($G$ is normalized such that the columns of $G$ sum 1). The intuitive idea is that in general, there exists some correlation between a person's interest. If one user likes sports, then he is likely to be interested in Health \& Fitness. This kind of information may help us to predict users' interest more accurately, especially for users with very few images in their potential interest categories. Therefore, during the label propagation, we also consider the propagation of category-level information.

To incorporate the group level information into our model, we define the following iterative process for image label propagation.
\begin{equation}
Y^{t+1} = ( 1 - \Lambda) WY^t G+ \Lambda Y^0
\label{eqn:lp}
\end{equation}
where $Y^0$ is the initial prediction of image labels from the trained CNN model, $W$ is the normalized similarity matrix between images and $\Lambda$ is a diagonal matrix. Following Yu et al.~\shortcite{yu2011collection}), $\Lambda$ is defined as
\begin{equation}
\Lambda_{i,i} = \max_{j}\frac{Y^0_{i,j}}{\sum_k{Y^0_{i,k}}}.
\label{eqn:lambda}
\end{equation}
We can consider the above process as two stages. In the first stage $( 1 - \Lambda) WY^t$, we use the similarity between the images to propagate the labels at image level. In the second stage, the group relationship matrix $G$ is employed to further propagate the group relationship to all images, i.e. we multiply matrix $G$ with the results of $( 1 - \Lambda) WY^t$. Next, we present the convergence analysis of the proposed label propagation framework.
\begin{algorithm}[h!]
\caption{User Profiling by Group Constraint Label Propagation}
\label{alg:glp}
\begin{algorithmic}[1]
\REQUIRE $X=\{x_1,x_2,\dots,x_N\}$ a collection of images\\
\quad \quad \ \  $M$: Fine-tuned Imagenet CNN model \\
\quad \quad \ \  $G$: Correlation matrix between the labels
\STATE Predict the categories $Y^0\in R^{N\times K}$ ($K$ is the number of categories) of $X$ using the trained CNN model $M$
\STATE Extract deep features from $M$ for all $X$.
\STATE Calculate the similarity matrix $W'$ between $X$ using Gaussian kernel function $W'(i,j) = \exp(-\frac{\parallel x_i - x_j\parallel^2}{2\delta^2})$, where $x_i$ and $x_j$ are the deep features for image $i$ and $j$ respectively. \label{alg:state:w}
\STATE Normalize $W'$ to get $W=D^{-1}W'$, where $D$ is diagonal matrix with $D_{ii} = \sum_j W'_{i,j}$.
\STATE Calculate the diagonal matrix $\Lambda$ according to Eqn.(\ref{eqn:lambda}).
\STATE Calculate the affinity matrix between $G$ between different categories\footnotemark.
\STATE Initialize, iteration index $t = 0$
\REPEAT
    \STATE Employ Eqn.(\ref{eqn:lp}) to update $Y^{t+1}$ according to $Y^t$
\UNTIL{Convergence or $t$ reaches the maximum iteration number}
\STATE Normalize rows of $Y^t\in R^{N \times K}$ to get $Y'^{t}\in R^{N \times K}$.
\RETURN $Y'^{t}$.
\end{algorithmic}
\end{algorithm}
\footnotetext{In our implementation, we use Jaccard index to calculate $G$, see the experimental section for detailed description.}
\subsubsection{Convergence Analysis}
From Eqn.~(\ref{eqn:lp}), we have the following formula for $Y^{t+1}$.
\begin{equation}
Y^{t+1} = \left(\left(1 - \Lambda\right)W\right)^{t+1} Y^0 G^{t+1} + \sum^{t}_{i = 0} \left(\left(1- \Lambda\right)W\right)^i\Lambda Y^0 G^i
\end{equation}
Since we have $0 \leq \lambda_{i,j}, w_{i,j}, G_{i,j} < 1$ for all $i,j$, therefore $\lim_{t \rightarrow \infty} \left(\left(1 - \Lambda\right)W\right)^{t+1} Y^0 G^{t+1} = 0$.
It follows that
\begin{equation}
\lim_{t \rightarrow \infty} Y^{t} =   \sum^{t-1}_{i = 0} \left(\left(1- \Lambda\right)W\right)^i\Lambda Y^0 G^i .
\end{equation}
According to the theorem that \textit{the product of two converging sequences converges}~\cite{rosenlicht1968introduction}, we define two different matrix series $A^{t} = \left(\left(1- \Lambda\right)W\right)^t\Lambda Y^0$ and $B^{t} = G^t$.
For matrix series $A^t$, it converges~\cite{yu2011collection}. For matrix series $B^t$, it also converges since $\rho (B) < 1$. Therefore, we conclude that the product of the two matrix series $A^t$ and $B^t$ also converges. More importantly, we have $\sum_{i} A_i  = \left( I - \Lambda\right) \Lambda Y^0 $ and $\sum_{i} B_i = (I - Q)^{-1}$. All the elements of $A_i$ and $B_i$ are non-negative, which leads to the fact that $\sum_i {A_iB_i} <_p (\sum_i A_i)(\sum_i B_i)$. We use $<_p$ to represent the element-wise comparison between two matrix. In other words, we have a upper bound for $Y^t$, together with the fact that $A_iB_i \ge 0$, we conclude that $Y^t$ converges.

Meanwhile, since we implement label propagation in each user's collection of images, thus the main computational and storage complexity is in the order of $O(n_i^2)$, where $n_i$ is the number of images for user $i$.

Algorithm~\ref{alg:glp} summarizes the main steps using the proposed group constrained label propagation framework. Note that in step~\ref{alg:state:w}, we use a Gaussian kernel function to calculate the similarity between two images using the deep features. However, one can employ other techniques such as locally linear embedding (LLE)~\cite{donoho2003hessian} to calculate the similarity between different instances.

Eventually, we obtain users' interests distribution by aggregating the label distribution of the collections of their images. In other words, we simply sum up the label distribution of individual images and then normalize it to produce the users' interest prediction results.

\section{Experiments and Evaluations}
To evaluate the proposed algorithm, we crawl data from Pinterest according to a randomly chosen user list consisting of 748 users. \tablename~\ref{tab:data:stat} gives the statistics of our crawled dataset. This dataset is used to train our own deep convolutional neural network. We use $80\%$ of these images as the training data and the remaining $20\%$ as validation data. We also downloaded another $77$ users' data as our testing data. Since each pinboard has one category, we use the category label as the label for all the images in that pinboard.
\begin{table}[thb]
    \vspace{-10pt}
    \caption{Statistics of our dataset.}
    \label{tab:data:stat}
    \vspace{5pt}
    \centering
    \begin{tabular}{|l|l|l|}
    \hline
    & Training and Validating & Testing \\ \hline
%    Num of Users & 748 & 84 \\ \hline
%    Num of Pinboards & 30,213 & 1,417 \\ \hline
%    Num of Pins & 1,586,947 & 158,406\\ \hline
    Num of Users & 748 & 77 \\ \hline
    Num of Pinboards & 30,213 & 1,126 \\ \hline
    Num of Pins & 1,586,947 & 66,050\\ \hline
    \end{tabular}
\end{table}

We train the CNN on top of the ImageNet convolutional neural model~\cite{krizhevsky2012imagenet}. We use the publicly available implementation Caffe~\cite{Jia13caffe} to train our model. All of our experiments are evaluated on a Linux X86\_64 machine with 32G RAM and two NVIDIA GTX Titan GPUs. We finish the training of CNN after a total of $200,000$ iterations.

\begin{figure}
\includegraphics[width=0.45\textwidth]{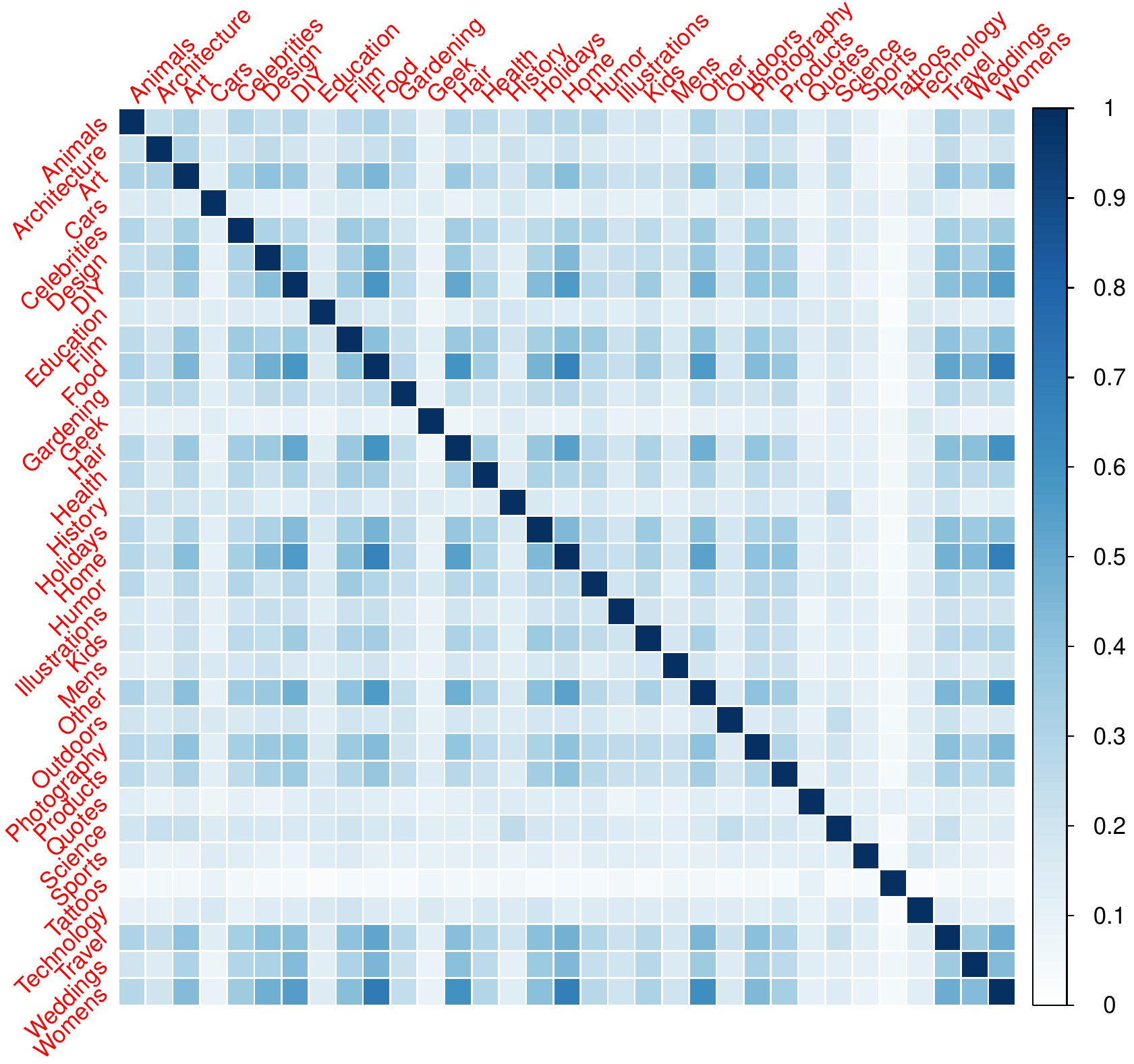}
\caption{Jaccard similarity coefficients between different categories in~\tablename~\ref{tab:pin:cat}.}
\label{fig:group}
\vspace{-15pt}
\end{figure}

In the proposed group constrained label propagation, we need to know the similarity matrix $G$ between all categories. We propose to learn the similarity matrix $G$ from pinterest users' behaviors of building their collections of pinboards. No additional textual semantic or visual semantic information is employed to build the similarity matrix. In our experiments, we employ Jaccard index to calculate the similarity coefficient between different categories in~\tablename~\ref{tab:pin:cat}. Specifically, the entry of $G_{ij}$ is defined as the Jaccard index between category $i$ and $j$, which is the ratio between the number of users who have pinboards of both categories $i$ and $j$ and the number of users who have pinboards of category $i$ or $j$.~\figurename~\ref{fig:group} shows the coefficients between all different categories. It shows that users prefer to choose some of the categories together, which may suggest potential category recommendations for Pinterest users.

\subsection{Evaluation Criteria}
To evaluate the performance of different models, we employ two different criteria. 1) \textbf{Normalized Discounted Cumulative Gain (NDCG) score} NDCG is a popular measure for ranking tasks~\cite{croft2010search}. Discounted Cumulative Gain at position $n$ is defined as
\begin{equation}
DCG_n = p_1 + \sum_{i=1}^n \frac{p_i}{\log_2(i)},
\end{equation}
where $p_i$is the relevance score at position $i$. In our case, we use the predicted probability as the relevance score. NDCG is the normalized DCG, where the value is between $0$ and $1$.
To calculate the NDCG, for a user $u_i$, we first get the ground truth distribution of his interest according to the category labels of his pinboards and then rank the ground truth categories in descending order of their distribution probabilities. Note that the pinboard labels of $u_i$ may not contain all the categories and our trained classifier may give a class distribution over all the categories. Therefore, when calculating the NDCG score we add a prior to the interest distribution of each user. We first calculate the prior category distribution $p_0$ from all the training and validating data set according to their labels. Then, for each testing user $i$ we first calculate the category distribution of $p_i$ according to the ground truth labels. Next, we smooth the distribution by updating $p_i = p_i + 0.1*p_0$ and then normalize it to get the ground truth interest distribution for user $i$. In this way, we are able to calculate the NDCG score for each user according to the prediction of the model. 2) \textbf{Recall@Rank K} In this metric, we firstly build the ground truth for one user $u_i$, we rank all of his categories in descending order according to the distribution probabilities. Then, we calculate the shared top ranked categories with the ground truth. Users' data in Pinterest may be incomplete. For example, user $i$ may be interested in Sports and Women's Fashion. However, she only has created pinboards belonging to Women's Fashion. She has not created Sports related pinboards. However, it does not mean that she is not interested in Sports. Thus, it makes sense to evaluate different algorithms' performance using the Recall@Rank k.
\begin{figure}[htbp]
\centering
\includegraphics[width=.4\textwidth]{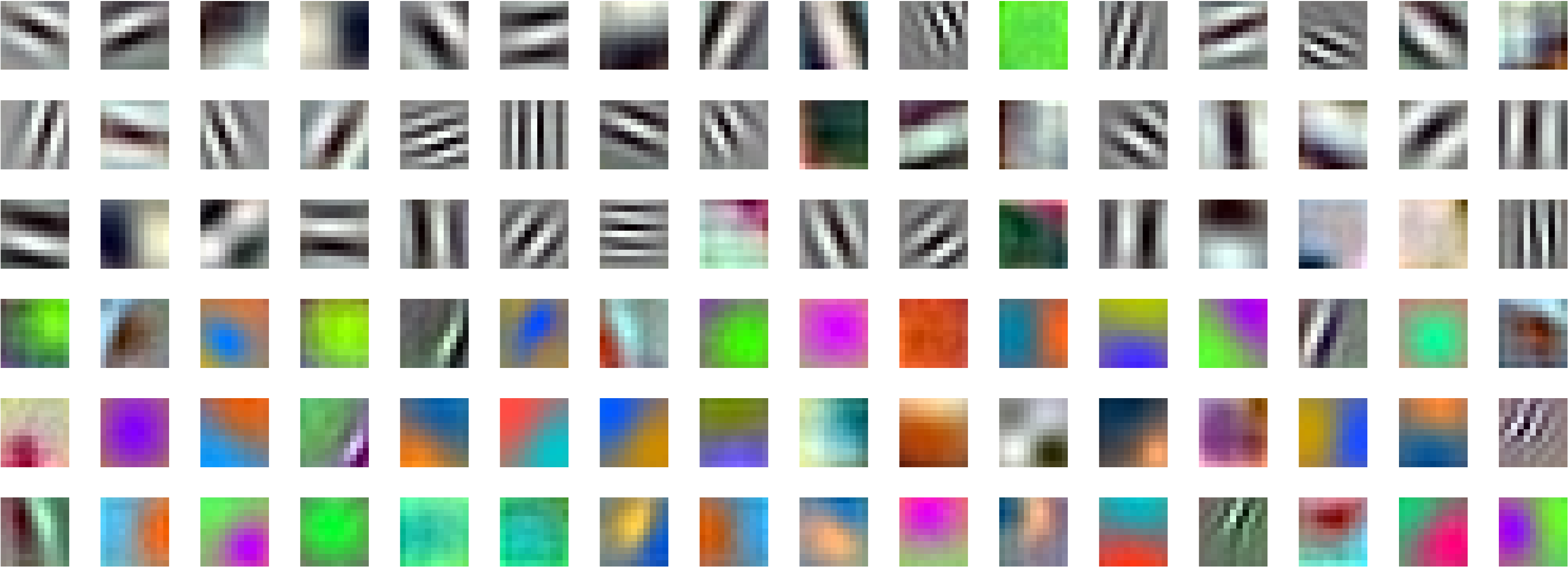}
\caption{Filters of the first convolutional layer.}
\label{fig:filters}
\vspace{-10pt}
\end{figure}
\subsection{Experimental Results}
%\begin{figure*}[!t]
%\begin{centering}
%\subfigure[Distribution of NDCG Score]{
%        %\begin{minipage}{.3\textwidth}
%            \centering
%            %\epsfig{file=./figures/upd.eps,width=0.45\textwidth}
%            \includegraphics[width=.3\textwidth]{./figure/pinterest_test_ndcg_score_full_soft}
%        %\end{minipage}
%         \label{fig:ndcg:cnn}
%}
%\subfigure[Distribution of NDCG Score with LP]{
%        %\begin{minipage}{.3\textwidth}
%            \centering
%            %\epsfig{file=./figures/upd.eps,width=0.45\textwidth}
%            \includegraphics[width=.3\textwidth]{./figure/pinterest_test_ndcg_score_lp_cnn_full_soft}
%        %\end{minipage}
%         \label{fig:ndcg:cnn:lp}
%}
%\subfigure[Distribution of NDCG Score with GLP]{
%        %\begin{minipage}{.3\textwidth}
%            \centering
%            %\epsfig{file=./figures/upd.eps,width=0.45\textwidth}
%            \includegraphics[width=.3\textwidth]{./figure/pinterest_test_ndcg_score_lp_group_cnn_full_soft}
%        %\end{minipage}
%         \label{fig:ndcg:cnn:glp}
%}
%\caption{Distribution of NDCG scores for three different approaches.}
%\label{fig:ndcg}
%\end{centering}
%\end{figure*}
\begin{figure}[htbp]
\centering
\includegraphics[width=.45\textwidth]{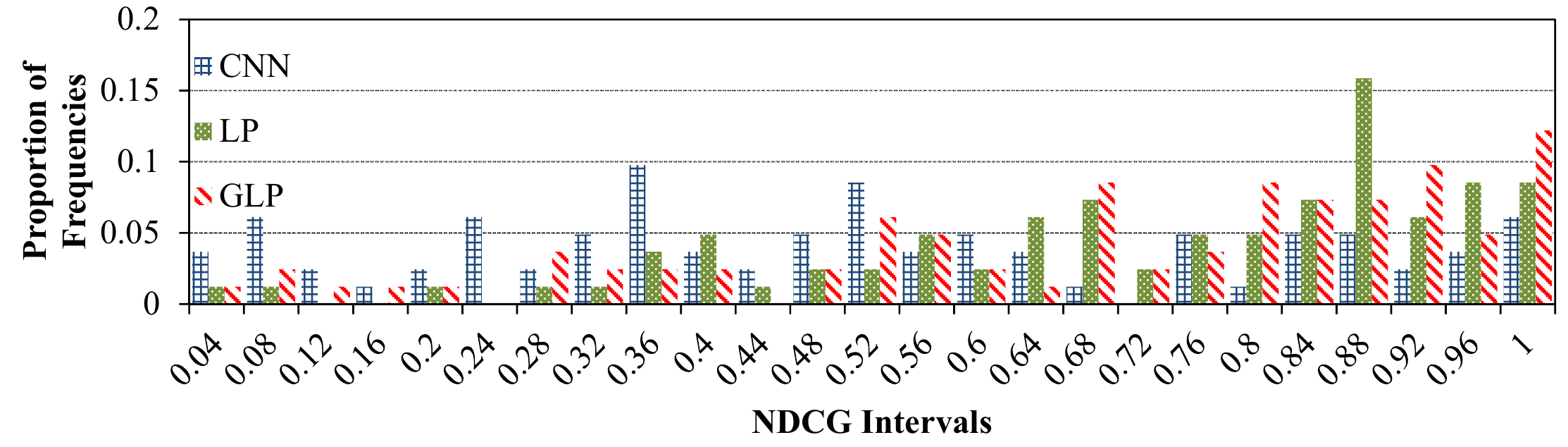}
\vspace{-5pt}
\caption{Distribution of NDCG scores for three different approaches.}
\label{fig:ndcg}
\vspace{-10pt}
\end{figure}
We report the results for three different models. First one is the results from our fine-tuned CNN model, i.e. the initial prediction for label propagation. Second model is label propagation~\cite{yu2011collection} (LP) and the last one is the proposed group constrained label propagation (GLP).

We first train our CNN model using the data in~\tablename~\ref{tab:data:stat}. \figurename~\ref{fig:filters} shows the learned filters in the first layer of the CNN model, which is consistent with the filters learned in other image related deep learning tasks~\cite{zeiler2013visualizing}. These Gabor-like filters can be employed to replicate the properties of human brain V1 area~\cite{lee2008sparse}, which can detect low-level texture features from raw input images. Next, we employ this model to predict the labels of the testing data, which is going to be the initial label prediction for LP and GLP models.
\begin{table}[!thb]
    \vspace{-15pt}
    \caption{Mean and Standard deviation of NDCG score}
    \vspace{5pt}
    \label{tab:ndcg:mean:std}
    \centering
    \begin{tabular}{|l|l|l|}
    \hline
%    Model & Mean & STD \\ \hline
%    CNN &0.6918 & 0.1785 \\ \hline
%    LP & 0.8187  & 0.1382\\ \hline
%    GLP & 0.8257 & 0.1380\\
    Model & Mean & STD \\ \hline
    CNN &0.692 & 0.179 \\ \hline
    LP & 0.818  & 0.138\\ \hline
    GLP & 0.826 & 0.138\\
    \hline
    \end{tabular}
    \vspace{-10pt}
\end{table}

For both LP and GLP models, we choose the $\delta$ in the Gaussian kernel to be the variance of distance between each pair of deep features~\cite{von2007tutorial}. We set the maximum number of iterations to be $100$ for both models\footnote{In our experiments, the increase of iteration number over 100 leads to similar results with the reported results in this work.}.
\figurename~\ref{fig:ndcg} shows the performance of using NDCG for the three models. The results shows that both LP and GLP try to move the distribution to the right, which suggests that both models can improve the performance in terms of NDCG score for most of the users. To quantitatively analyze the results, we give the mean and the standard derivation of the NDCG scores in~\tablename~\ref{tab:ndcg:mean:std}. Both LP and GLP improve the NDCG score from about $0.69$ to over $0.80$ and reduce the standard deviation. GLP shows slight advantage over LP in terms of both mean and standard deviation of NDCG score.

\begin{figure}[htbp]
\centering
\includegraphics[width=.45\textwidth]{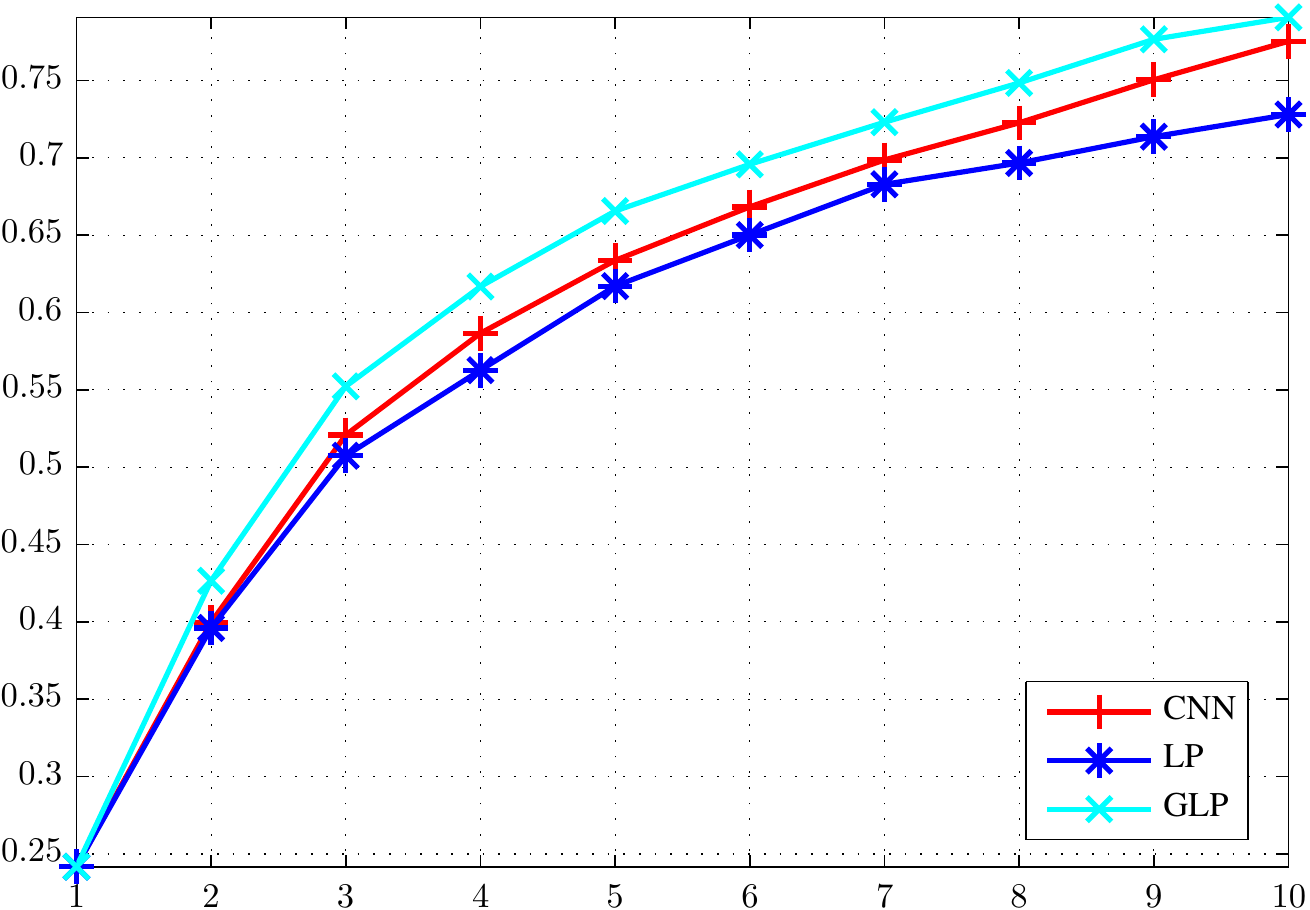}
\caption{Recall@$K$ for different $K$ for the three models respectively.}
\label{fig:recall}
\vspace{-10pt}
\end{figure}
The evaluation results of using Recall@K are shown in~\figurename~\ref{fig:recall}. Differently from the NDCG score, CNN and GLP show better recall performance then LP for all different values of $K$s. Meanwhile, GLP consistently outperforms the CNN model. The results suggest that without the propagation of category or group similarity, label propagation may cause the increase of irrelevant categories, which leads to poor recall.

Besides the user-level interests prediction, we are also interested in \textit{whether or not label propagation can improve the performance of image-level classification performance}. Since we have the pinboard labels as the ground-truth labels for each testing image, we report the accuracy of the three models on the testing images.~\tablename~\ref{tab:img:ac} summarizes the accuracy. Indeed, the performance of these three models is quite similar. These results may suggest that the label propagation have limited impact on the overall label distribution of individual image, i.e. the max index in the probability distribution of labels. However, they may change the probability distribution, which may benefits the overall user-level interests estimation.
%~\tablename~\ref{tab:img:ac} summarizes the accuracy. Interestingly, the image-level performance of both LP and GLP outperforms the performance of CNN. Meanwhile, even though LP gives better performance than GLP in this image-level evaluation, it fails to give the better performance than GLP in the user-level interests prediction in terms of both NDCG score and Recall@K. These results help us to conclude that the propagation of group correlation helps not only in image-level classification, but also collection-level or user-level classification.
\begin{table}[!thb]
    \vspace{-15pt}
    \caption{Accuracy on Image-level classification of different models.}
    \label{tab:img:ac}
    \vspace{5pt}
    \centering
    \begin{tabular}{|l|l|}
    \hline
%    Model & Accuracy \\ \hline
%    CNN & 0.4176 \\ \hline
%    LP & 0.4333 \\ \hline
%    GLP & 0.4204 \\
%    Model & Accuracy \\ \hline
%    CNN & 0.418 \\ \hline
%    LP & 0.433 \\ \hline
%    GLP & 0.420 \\
    Model & Accuracy \\ \hline
    CNN & 0.431 \\ \hline
    LP & 0.434 \\ \hline
    GLP & 0.434 \\
    \hline
    \end{tabular}
    \vspace{-10pt}
\end{table}

\section{Conclusions}
We addressed the problem of user interest distribution by analyzing user generated visual content. We framed the problem as an image classification problem and trained a CNN model on training images spanning over 748 users' photo albums. Taking advantage of the human  intelligence incorporated through the user curation of the organized visual content, we used image-level similarity to propagate the label information between images, as well as utilized the image category information derived from the user created organization structure to further propagate the category-level knowledge for all images. Experimental evaluation using data derived from Pinterest provided support for the effectiveness of the proposed method. In this work, our focus was on using image content for user interest prediction. We plan to extend our work to also incorporate user generated textual content in the form of comments and user profile information for improved performance.

\begin{small}
\bibliographystyle{named}
\bibliography{aaai_2015}
\end{small}

\end{document}